\documentclass[prx,aps,twocolumn,pdflatex,showpacs,superscriptaddress]{revtex4}
\usepackage{amsmath, amsthm, amssymb,float}
\usepackage{amsfonts}
\usepackage{graphicx}
\usepackage{dcolumn}
\usepackage{bm}
\usepackage{textcomp}
\usepackage[normalem]{ulem}
\usepackage{mathrsfs}
\usepackage{color}
\newcommand{\bl}[1]{{\color{blue}#1}}

\usepackage{epsfig}
\usepackage{textcomp}
\usepackage{epstopdf}

\begin{document}

\title{Generation of High Order Harmonics  in Heisenberg-Euler Electrodynamics}

\author{P. V. Sasorov}
%\email{pavel.sasorov@eli-beams.eu}
\affiliation{Institute of Physics of the ASCR, ELI--Beamlines project, Na Slovance 2, 18221, Prague, Czech Republic}
\affiliation{Keldysh Institute of Applied Mathematics, Moscow, 125047, Russia}
\author{F. Pegoraro}
\affiliation{Enrico Fermi Department of Physics, University of Pisa, Italy
and National Research Council, National Institute of Optics, via G. Moruzzi 1, Pisa, Italy}
\author{T. Zh. Esirkepov}
\affiliation{National Institutes for Quantum and Radiological Science and Technology (QST),
Kansai Photon Science Institute, 8--1--7 Umemidai, Kizugawa, Kyoto 619--0215, Japan}
\author{S.  V. Bulanov}
%\email{sergei.bulanov@eli-beams.eu}
\affiliation{Institute of Physics of the ASCR, ELI--Beamlines project, Na Slovance 2, 18221, Prague, Czech Republic}
\affiliation{National Institutes for Quantum and Radiological Science and Technology (QST),
Kansai Photon Science Institute, 8--1--7 Umemidai, Kizugawa, Kyoto 619--0215, Japan}

\date{\today}

\begin{abstract}

High order harmonic generation by  extremely intense, interacting,  electromagnetic waves in the quantum  vacuum 
is investigated within the framework of the Heisenberg-Euler formalism. Two intersecting plane waves of finite duration 
are considered in the case of general polarizations. Detailed finite expressions are obtained for the case 
where only the first Poincar\'e invariant does not vanish. Yields of high harmonics in this case are most effective.

\end{abstract}

\maketitle

\tableofcontents
\nopagebreak

\section{Introduction}
\label{sec:intro}

Present and forthcoming high power laser developments \cite{D2019, YP21, LI21} open a vast
area in the study of nonlinear physics phenomena related to the electromagnetic field interaction with matter and
vacuum~\cite{WG1985, DG00, Mo06,Ma06,TG09,DiP12, ZHANG20, SEI20, ROB21}. 
Vacuum polarization effects leading to 
high order harmonics generation in vacuum
in the course of  the collision of the extreme intensity 
laser pulses is one of the more important directions in these investigations \cite{AAM85, KING16, AB19, TMJ20}.  
This problem has attracted substantial attention 
\cite{NNR93,FG15, Ka00,VB80,SHI20,P05,Lu06,Nar07,Fed07, AP14, Bohl15, FK19, SH19}
because it sheds  light on  the dynamical properties of Quantum Electrodynamics (QED) vacuum 
in strong electromagnetic  fields. The  generation of high order harmonics plays an important role in the steepening 
of  a nonlinear electromagnetic wave,
in  the intersecting of strong laser beams ~\cite{Ka19}   and  in the formation of the relativistic electromagnetic solitons 
in the QED vacuum~\cite{Sol19, FP21}.

In spite of a number of publications \cite{Ka00, P05,Lu06,Nar07,AP14,Fed07, Bohl15} 
devoted to the detailed theoretical analysis of the harmonic generation,
 the theory of this process is far from complete. 
 There are several theoretical questions that need to be more thoroughly clarified. 
 They concern the role of the  various  small parameters governing the problem 
  that should be ordered and considered simultaneously, 
taking into account a specific form of the Heisenberg-Euler Lagrangian  \cite{H-E, W36, ADS, BLP} 
describing the QED vacuum polarization effects.

The present paper is devoted to  the theoretical consideration and clarification of  the high order harmonics 
generation mechanism 
within the framework of the Heisenberg-Euler electrodynamics. We consider here the case where the high order harmonics 
are generated by two  crossing  plane electromagnetic waves in vacuum under the conditions where
only one of  the two Poincar\'e invariants 
of the electromagnetic field  do not vanish. We assume that the finite length electromagnetic 
pulses are infinite in the two transverse directions. 
Applying a  Lorentz transformation  we can find a boosted frame of reference where the crossing 
electromagnetic pulses appear 
in the form  of two counter-propagating waves (e.g. see \cite{KIRK}). Thus the problem becomes 
one-dimensional,  non-stationary  and is  described by two independent variables. 
We consider the case of electromagnetic waves with  wavelengths much larger than  the electron 
Compton scattering length and with  amplitudes substantially 
lower than the Schwinger field, which corresponds to the conditions of validity of the 
Heisenberg--Euler approach~\cite{H-E, W36, ADS, BLP}. 
 In this context we recall  that the problem of extending the Heisenberg--Euler electrodynamics 
 has attracted much interest from the quantum field theory community. One of the ways along which  
 such a generalization can be performed  is  by taking into account  the two-loop correction to the ground 
 state energy of the virtual electron-positron sea in an almost constant external electromagnetic field, see for 
 example \cite{GK17} and references therein. Here  we prefer  to consider only  the lowest non-vanishing order 
 in terms of powers of fine-structure constant $\alpha$, assuming  the amplitudes of in-coming waves in terms 
 of the Schwinger field as given. This approach allows us to  restrict our investigation  to  the original  
 Heisenberg--Euler Lagrangian  and to simplify considerably our results in the framework of the Heisenberg--Euler electrodynamics approach.

The paper is organized as  follows. In Section~\ref{Vacuum} we present the nonlinear wave equations
within the framework of the Heisenberg--Euler electrodynamics 
which are used throughout  the paper. Section~\ref{def} describes the electromagnetic configuration and the 
Dirac light-cone coordinates convenient for analysing  the problem under consideration. Section~\ref{currcons} is devoted 
to the derivation of a convenient form of nonlinear wave equation. The formulation of the scattering problem 
and  the introduction 
of a perturbation theory which  is used for obtaining final results are in Section \ref{initcon}. In Section~\ref{harm} 
we write the expressions 
giving the intensity of the high order harmonics.  Section~\ref{APP} is devoted to a general 
case corresponding to different polarizations. The discussion and  the summary of the results obtained 
are presented in Section~\ref{concl}.

\section{Nonlinear electrodynamics equations describing the quantum vacuum}
\label{Vacuum}

The  analysis of the electromagnetic wave interaction in the QED vacuum 
 is based on the Heisenberg--Euler  Lagrangian density, $\mathcal{L}_{HE}$ \cite{H-E, W36, ADS, BLP}.  
The sum of the Lagrangians, 
\begin{equation}
\mathcal{L}=\mathcal{L}_{0}+\mathcal{L}_{HE}, \label{eq:Lagrangian}
\end{equation}
describes
the electromagnetic field in the long-wavelength limit. Here
\begin{equation}
\mathcal{L}_{0}=-\frac{m^4}{16\pi\alpha}F_{\mu \nu}F^{\mu \nu}
\end{equation}
 is the  Lagrangian of classical electrodynamics with  the electromagnetic field tensor $F_{\mu \nu}$ 
 defined  in terms of the 4-vector potential $A_\mu$  as \cite{FT}
 \begin{equation}
\label{eq:Fmn}
F_{\mu \nu}=\partial_{\mu} A_{\nu}-\partial_{\mu} A_{\nu}.
\end{equation}
Here and in the following we use dimensionless variables with  $\hbar=c=1$, $m_e$ and $e$ are  the electron mass   
and elementary electric 
charge, and $\alpha = e^2/(4\pi)\approx 1/137$ is the fine structure constant.  Thus the electromagnetic fields  are measured 
in units of $m_e^2/e$ , i.e.  are normalized on the QED critical field  $E_S = m_e^2 c^3/{e\hbar} = 1.32 \times 10^{18} V/m$.

In the Heisenberg--Euler theory, the radiative corrections are described by the $\mathcal{L}_{HE}$ 
term on the right hand side of Eq.(\ref{eq:Lagrangian}).
It can be written as \cite{BLP}
\begin{align}
\label{eq:HELagr} &
\mathcal{L}_{HE}=\frac{m_e^4} {8 \pi^2}{\cal M}( {\mathfrak e}, {\mathfrak b})= 
\frac{m^4}{8\pi^2}\int^{\infty}_0 \frac{\exp{(-\eta)}}{\eta^3}\times &\\
&\left[-(\eta {\mathfrak e}\cot \eta  {\mathfrak e}) (\eta {\mathfrak b}\coth \eta  {\mathfrak b}) 
+1-\frac{\eta^2}{3}({\mathfrak e}^2-{\mathfrak b}^2)\right] d \eta.\nonumber
\end{align}
The invariant fields  $ {\mathfrak e}$ and $ {\mathfrak b}$
are expressed in terms of the Poincar\'e invariants
\begin{equation}
\label{eq:FGinv}
{\mathfrak F}=\frac{1}{4}F_{\mu \nu}F^{\mu \nu}
\quad {\rm and} \quad
{\mathfrak G}={\frac{1}{4}F_{\mu \nu}\tilde F^{\mu \nu}}
\end{equation}
 as
\begin{equation}
\label{eq:abinv}
{\mathfrak e}=\sqrt{\sqrt{{\mathfrak F}^2+{\mathfrak G}^2}-{\mathfrak F}} \,\,\, {\rm and}  \,\,\, {\mathfrak b}
=\sqrt{\sqrt{{\mathfrak F}^2+{\mathfrak G}^2}+{\mathfrak F}},
\end{equation}
respectively. The dual tensor $\tilde F^{\mu \nu}$ is defined by
\begin{equation}
\tilde F^{\mu \nu}= {\frac{1}{2}}\varepsilon^{\mu \nu \rho \sigma}F_{\rho \sigma}\, ,
\end{equation}
where  $\varepsilon^{\mu \nu \rho \sigma}$ is the Levi-Civita symbol in four dimensions
{($\varepsilon^{0123}=-\varepsilon_{0123}=-1$)}.
In the 3D notation the Poincar\'e invariants are
\begin{equation}
\mathfrak{F}=\frac{1}{2}\left({\bf B}^2-{\bf E}^2\right),\quad
\mathfrak{G}={\bf B}\cdot{\bf E}.
\label{3d}
\end{equation}

As explained in Ref.~\cite{BLP},
the  Heisenberg--Euler Lagrangian in the form given by Eq.(\ref{eq:HELagr})
  should be used for obtaining an asymptotic
series over the field invariant
 $ {\mathfrak e}$ and ${\mathfrak b}$ assuming that  they are small.
In this limit, the function ${\cal M}( {\mathfrak e}, {\mathfrak b})$ in Eq.(\ref{eq:HELagr}) can be expanded
for  small arguments as:
\begin{align}
\label{eq:M}
&{\cal M}({\mathfrak e},{\mathfrak b})=\frac{\Gamma(2)}{45}
\left[({\mathfrak e}^4+{\mathfrak b}^4)+5 {\mathfrak e}^2\,{\mathfrak b}^2\right]\nonumber \\
&-\frac{\Gamma(4)}{945}
\left[2({\mathfrak b}^6-{\mathfrak e}^6)+7{\mathfrak e}^2\,{\mathfrak b}^2({\mathfrak b}^2-
{\mathfrak e}^2)\right]\\
&+\frac{\Gamma (6)}{7\times 45^2}\left[3\left({\mathfrak e}^8+{\mathfrak b}^8\right)
+10{\mathfrak e}^2{\mathfrak b}^2\left({\mathfrak e}^4+{\mathfrak b}^4\right)-7{\mathfrak e}^4{\mathfrak b}^4\right]
+\dots \nonumber \\
\nonumber
\end{align}
Here $\Gamma(x)$ is the Euler Gamma function \cite{DLMF}.
The properties of the expansion of ${\cal M}({\mathfrak e},{\mathfrak b})$ in series of powers 
of the fields $\mathfrak e$ and  $\mathfrak b$ are discussed in Ref.~\cite{Dunne}.
The expression (\ref{eq:M}) yields for the Lagrangian $\mathcal{L}_{HE}$
in the weak field approximation  (see also Ref. \cite{HeHe})
\begin{align} 
& \mathcal{L}_{HE}=\kappa {\cal M}\!\!\!\!\!&=&\frac{\kappa}{45}\left(4 {\mathfrak F}^2
+ 7 {\mathfrak G}^2\right) -\frac{4\kappa}{315} {\mathfrak F}
\left(  8 {\mathfrak F}^2 +13 {\mathfrak G}^2  \right)
\nonumber\\ &
&+&\frac{8\kappa}{945}\left(48\mathfrak{F}^4+88\mathfrak{F}^2\mathfrak{G}^2
+19\mathfrak{G}^4\right)+\dots
\label{eq:mathcalL}
\end{align}
with the constant
$\kappa= {m}^4/8 \pi^2$.
In the Lagrangian given by Eq.(\ref{eq:mathcalL}) the first, second and third terms on the right hand side 
correspond  to  four-, six- and eight-photon interactions, respectively.

\section{The electromagnetic field configuration in the Dirac light-cone coordinates }\label{def}

As explained in the introduction we consider  two counter propagating waves along the  $x$-axis. 
The vector potential corresponding to the electromagnetic field  can be presented in the form
\begin{equation}\label{VP}
\mathbf{A}=A_z(t,x) \mathbf{e}_z + A_y(t,x) \mathbf{e}_y.
\end{equation}

Below we use  Dirac's
light cone coordinates ${x^-}$ and ${x^+}$ defined as (see  e.g. Ref.~\cite{LCD})
\begin{equation}
\label{eq:x+x-}
x^+=\frac{x+t}{\sqrt{2}},\quad
x^-=\frac{t-x}{\sqrt{2}}.
\end{equation}

The Lorentz transform  of the light-cone variables  under a boost along $x$ with velocity $\beta$
is given by
\begin{align}
\label{Lor1}
& x^{\prime +}=\frac{x'+t'}{\sqrt{2}}=e^{- \eta} \frac{x+t}{\sqrt{2}}=e^{- \eta}x^+,  \\
&x^{\prime -}=\frac{t'-x'}{\sqrt{2}}=e^{ +\eta} \frac{t-x}{\sqrt{2}}=e^{+ \eta}x^- ,
\end{align}
with  a prime denoting  variables in the boosted frame of reference and $\eta$ equal to
\begin{equation}
\label{eq:eta}
\eta ={\rm ln}\sqrt{\frac{1+\beta}{1-\beta}}.
\end{equation}

The following shorthand  stands for the partial derivatives:
\begin{equation}\label{der}
\partial_+f=\left(\frac{\partial f}{\partial x^+}\right)_{x^-}\, ,
\quad
\partial_-f=\left(\frac{\partial f}{\partial x^-}\right)_{x^+}\, .
\end{equation}
Then the derivatives are transformed as:
\begin{equation}\label{Lor2}
\left(\partial_-\right)^\prime=e^{- \eta}\partial_-\,
\quad {\rm and} \quad
\left(\partial_+\right)^\prime=e^{+ \eta}\partial_+\,  ,
\end{equation}
respectively.

For simplicity we consider in this section interacting electromagnetic waves of the same polarization. 
The general polarization case is considered 
in Section~\ref{APP}.  In the case of the same polarizations, the invariant ${\mathfrak G}$ vanishes identically. The 
coordinate system can be chosen so that 
$A_y =0$. We will use a notation  $a(x,t)$ for $A_z(x,t)$.

Calculating the derivatives of $a(x,t)$ with respect to the coordinates $x^{ +}$ and $x^{-}$, 
\begin{equation}
\label{fields}
u=\partial_{-}a \quad {\rm and} \quad w=\partial_{+}a,
\end{equation}
we find the relationships between $u$ and $w$ and the electric, $e_z=-\partial_t a$,  and magnetic, $b_y = - \partial_x a$,
field components. They read 
\begin{equation}
\label{fields2}
u =-\frac{e_z-b_y}{\sqrt{2}} \quad {\rm and} \quad w  = -\frac{e_z+b_y}{\sqrt{2}}.
\end{equation}
The field variables $u$ and $w$ are not independent of each other. 
Due to commutation of partials, $\partial_{-+}a=\partial_{+-}a$, the field variables $u$ and $w$
obey the equation
\begin{equation}\label{constr}
\partial_-w=\partial_+u\, .
\end{equation}
The Lorentz transformation  of the fields $u$ and $w$ 
\begin{align}
\label{Lor3}
w^{\prime} &=-\frac{e_z^{\prime}+b_y^{\prime}}{\sqrt{2}}=-e^{+ \eta}\frac{e_z+b_y}{\sqrt{2}}=e^{+ \eta} w,\\
u^{\prime}& =-\frac{e_z^{\prime}-b_y^{\prime}}{\sqrt{2}}=-e^{- \eta}\frac{e_z-b_y}{\sqrt{2}}=e^{- \eta} u
\end{align}
shows that the $w$ and $u$ are transformed as the \bl{$-$ and $+$}  components of  a contravariant 4-vector, respectively.
The  field product $ uw = (b_y^2-e_z^2)/{2}$,
\begin{equation}
\label{eq:fields-uw}
u^{\prime} w^{\prime}=uw,
\end{equation}
is  Lorentz invariant and  is equal to the first Poincar\'e invariant $\mathfrak{F}$.

\section{Description of the wave interaction in terms of canonical momentum conservation equations} 
\label{currcons}

In the case  of interacting electromagnetic pulses with the same polarization, the Lagrangian given 
by Eq.(\ref{eq:Lagrangian}) 
can be written in terms of  the field variables $u$ and $w$, defined by Eq.(\ref{fields}),
as
\begin{equation}  \label{grangina1}
{\cal L}  = -\frac{m^4}{4\pi\alpha} [wu  +\alpha \, Q(wu)] =-\frac{m^4}{4\pi\alpha}\bar{\cal L}(u,w)\, .
\end{equation}
Here 
\begin{equation}  \label{barL}
\bar{\cal L}  =  wu  +\alpha \, Q(wu)
\end{equation}
is the normalized Lagrangian,
where it is assumed that  $uw\ll1$ and
\begin{equation}\label{Qdef}
Q(uw)=-\frac{1}{2\pi}{\cal M}\left(0,\sqrt{2uw}\right)\, .
\end{equation}
The function $Q(\zeta)$ can be represented in the form of  the series
\begin{equation}\label{Qser}
\quad Q(\zeta)=\sum\limits_{m=2}^\infty b_m \zeta^m\, 
\end{equation}
with  coefficients $b_m$
\begin{equation}\label{Qcoeff}
b_m=\frac{2^{3(m-1)} B_{2m}}{\pi m (2m-1)(m-1)}\, 
\end{equation}
proportional to the Bernoulli numbers~\cite{DLMF}, $B_n$. 
These coefficients can be obtained from the general expansion of ${\cal M}({\mathfrak e},{\mathfrak b})$ in series of 
powers of $\mathfrak e$, and  $\mathfrak b$,  (see Ref. \cite{Dunne}) setting ${\mathfrak e}=0$ and 
${\mathfrak b}=\sqrt{2\zeta}$, where $\zeta$ is the argument of the function $Q(\zeta)$ 
[see Eqs.~(\ref{Qdef}) and~(\ref{Qser})]. It represents actually the field invariant $\mathfrak{F}$ in 
Eq.~(\ref{3d}).
Several leading order coefficients $b_m$ in the expansion~(\ref{Qser}) are presented in Tab.~\ref{bn}. 
Using the asymptotic dependence of the Bernoulli numbers at  $m\to\infty$ (see Ref. \cite{DLMF} (Sec. 24.11))
\begin{equation}
B_{2m}\simeq (-1)^{m+1}4\sqrt{\pi m}\left(\frac{m}{\pi e}\right)^{2m}\, ,
\end{equation}
we obtain the asymptotic expression for the coefficients $b_m$, 
\begin{equation}
b_m\simeq \frac{(-1)^{m+1}}{4\sqrt{\pi m^5}}\left(\frac{2^{3/2}m}{\pi e}\right)^{2m}\, .
\end{equation}
We see that the coefficients of the series for $Q(\zeta)$ in Eq. (\ref{Qser}) grow faster than exponentially for  large $m$. 
Thus this series can be considered only as an asymptotic series and should be truncated by taking into account only  
a finite number $m_0$ of its terms. 
Nevertheless, this finite number $m_0$ is rather   large for the relevant values of $|uw|\ll 1$. The truncation number $m_0$ 
can be estimated as a maximum of $m$ for which the terms in Eq.~(\ref{Qser}) are  still decreasing. 
This criterion  gives
\begin{equation}
m_0\sim \frac{\pi e}{\sqrt{8 |uw|}}\, .
\end{equation}
Thus the condition $m\lesssim m_0$ is very weak allowing the use of a substantially large number 
of the terms in Eq.~(\ref{Qser}). 
This problem was considered  from different points of view also in Refs.~\cite{Dunne,HS06}.

\begin{table}[h]
\caption{Several coefficients in the series of $Q(\zeta)$, Eq.~(\ref{Qser})}
\label{bn}
\begin{tabular}{cc|cc}
\toprule
$m$&$b_m$& $m$&$b_m$\\
\colrule
&&\\
2&$\quad-\frac{2}{45\pi}$          &6&$\quad-\frac{5660672}{225225\pi}$   \\
&&\\
3&$\quad \frac{16}{315\pi}$        &7&$\quad\frac{65536}{117\pi}$        \\
&&\\
4&$\quad-\frac{64}{315\pi}$        &8&$\quad-\frac{474087424}{26775\pi}$   \\
&&\\
5&$\quad\frac{512}{297\pi}$         &9&$\quad\frac{45997883392}{61047\pi}$   \\
\end{tabular}
\end{table}

The Euler-Lagrange equation corresponding to the Lagrangian~(\ref{grangina1}) and the equation~(\ref{constr})
gives  the field equations.
They can be written as a current conservation equation
\begin{equation}
\partial_+ \left( \frac{\partial \bar{\cal L}}{\partial w }\right) + \partial_- \left( \frac{\partial \bar{\cal L}}{\partial u}\right)  = 0  .\label{cons.curr}
\end{equation}

We can define the canonical momenta
\begin{align}   \label{momentplus} &
\bar{\Pi}^+ = \frac{\partial \bar{\cal L}}{\partial w} = u[ 1 +\alpha Q'(wu)] ,  \\
 & \bar{\Pi}^-  =\frac{\partial \bar{\cal L}}{\partial u}  =  w[ 1 +\alpha Q'(wu)]  ,  \label{momentmin}
\end{align}
where $Q^\prime(\zeta)$ denotes differentiation of $Q(\zeta)$ with respect to  its single argument. 
Then Eq.~(\ref{cons.curr}) takes the form
\begin{equation}
\partial_+ \bar{\Pi}^+  \, + \, \partial_-  \bar{\Pi}^-  = 0  .\label{cons.moment}
\end{equation}

Using the above obtained relationships we can find the explicit form of the field equation. They can be written as
\begin{equation}\label{FE010}
\partial_+ [ u +u\alpha Q'(wu)]+\partial_- [ w +w\alpha Q'(wu)]=0\, ,
\end{equation}
\begin{equation}\label{FE020}
\partial_-w=\partial_+ u\, .
\end{equation}
We can use Eq.~(\ref{FE020}) to exclude either  $\partial_-w$  or $\partial_+u$ from Eq.~(\ref{FE010}).
 As a result we obtain a set of two  equivalent symmetric field equations:
\begin{align}
&\partial_+\left[u+\frac{\alpha}{2} u Q^\prime(uw)\right]=
-\frac{\alpha}{2}\partial_-\left[wQ^\prime(uw)\right],
\label{FE030}\\
&\partial_-\left[w+\frac{\alpha}{2} w Q^\prime(uw)\right]=
-\frac{\alpha}{2}\partial_+\left[uQ^\prime(uw)\right].
\label{FE040}
\end{align}
This form of the field equations is convenient for  applying  a perturbation approach because in the region
where $w=0$ Eq.~(\ref{FE030}) takes the form
\begin{equation}
\partial_+u=0\, ,\mbox{~~hence~~} u=u(x^-)\, ,
\end{equation}
whereas in the region where $u=0$  Eq.~(\ref{FE040}) becomes
\begin{equation}
\partial_-w=0\, ,\mbox{~~hence~~} w=w(x^+)\, .
\end{equation}
The momenta introduced above can be represented in the same regions as
\begin{equation}\label{FE050}
\bar{\Pi}^+=u=u(x^-)\, ,\quad \bar{\Pi}^-=0\,,
\end{equation}
\begin{equation}\label{FE054}
\bar{\Pi}^-=w=w(x^+)\, ,\quad \bar{\Pi}^+=0\,,
\end{equation}
respectively.

It is  also  useful to introduce the energy momentum tensor
\begin{equation}
\bar{T}_\lambda^\mu=a_\lambda\frac{\partial \bar{\cal L}}{\partial a_\mu}-\delta_\lambda^\mu\bar{\cal L}
\quad (\lambda\, , \mu =\pm)\, ,
\end{equation}
where $a_\lambda=\partial_\lambda a$ and $\delta_{\mu \nu}$ is the Kronecker delta.
The energy momentum tensor is useful to express the conservation of energy and momentum of the field
\begin{equation}
\partial_\mu T_\lambda^\mu =0.
\end{equation}

In terms of the variables $u$ and $w$ the tensor $\bar{T}_\lambda^\mu$ components can be written as 
\begin{eqnarray}
\bar{T}_+^+&=&w\frac{\partial \bar{\cal L}}{\partial w}-\bar{\cal L}=\alpha\left[ uwQ^\prime(uw)-Q(uw)\right],
\label{T010}\\
\bar{T}_-^+&=&u\frac{\partial \bar{\cal L}}{\partial w}=u^2\left[1+\alpha Q^\prime(uw)\right],
\label{T020}\\
\bar{T}_+^-&=&w\frac{\partial \bar{\cal L}}{\partial u}=w^2\left[1+\alpha Q^\prime(uw)\right],
\label{T030}\\
\bar{T}_-^-&=&u\frac{\partial \bar{\cal L}}{\partial u}-\bar{\cal L}=\alpha\left[ uwQ^\prime(uw)-Q(uw)\right].
\label{T040}
\end{eqnarray}
We see that the diagonal components, $\bar{T}_+^+=\bar{T}_-^-$, are equal to each other due the tensor symmetry $T_{tx}=T_{xt}$,  see also
Ref.~\cite{FP19}.

\section{Perturbation theory to obtain the solutions of the scattering problem}\label{initcon}

We formulate the scattering problem for the field equations~(\ref{FE010}-\ref{FE020}) or~(\ref{FE030}-\ref{FE040}).

In the case of non-interacting waves, the functions $u$ and $w$ do not change since they are equal to
$u=u(x^-)$ and $w=w(x^+)$.  
Let us consider the case where the functions  $u=u(x^-)$ and $w=w(x^+)$ describe the two pulses of  finite length. 
The length is $L_+$ for the pulse propagating in the 
$x^+$ direction and it equals $L_-$ in the case of the pulse propagating in the in the 
$x^-$ \(  \) direction. 
We assume that the amplitude of the $u(x^-)$ function is approximately equal to  $u_0$ at $|x^-|\lesssim L_-$, 
and it becomes exponentially small as compared with $u_0$ outside this region. 
We assume also that there are no spatial structures in the $u(x^-)$ pulse with scale-length much less than $L_-$. 
We set analogous assumptions for  the $w(x^+)$ function with the obvious substitution: $-\to +$. 
As  was noticed above, the pulse amplitudes are  substantially small, 
$u_0w_0\ll1$. 
We will consider the  case  where the nonlinear effects, leading to harmonics generation of different orders, 
can be treated perturbatively.  Thus, we should exclude too small and too large values of the ratio 
$L_-/L_+$. This point will be  considered below. 
Here we present only a simplified condition for the applicability of the  perturbation approach used below:
\begin{equation}\label{FE052}
\alpha w_0^2\ll \frac{L_-}{L_+}\ll \frac{1}{\alpha u_0^2}\, .
\end{equation}
This strong inequality is Lorentz-invariant, that means that  all its terms have the same dependence 
on the boost parameter $\eta$ given by Eq. (\ref{eq:eta}). 
This criterion will be made more precise below when we  will further specify the parameters of  the pulses.  

Let us now take into account the interaction between the waves $u$ and $w$. 
It takes place only in the following `rectangular' domain, $\Omega$, on the $(x^+,x^-)$-plane, where both the following  inequalities are valid:
\begin{equation}
|x^-|\lesssim L_-\,,\mbox{~~and~~}|x^+|\lesssim L_+\, .
\end{equation}
In this `rectangular' region $\Omega$, both amplitudes $u$ and $w$ depend on both  the  independent coordinates: 
$x^+$ and $x^-$. Outside this region, when $|x^-|\gg L_-$ or $|x^+|\gg L_+$, the waves do not interact with each other, 
and $w$ depends only on the coordinate $x^+$, whereas $u$ depends only on the coordinate $x^-$. 
We may distinguish {\it in}coming and {\it out}going waves in the latter combined region:
\begin{eqnarray}
& w(|x^+|\lesssim L_+\, , x^-\gg L_-)&=w^{in}(x^+)\, ,
\label{FE060}\\
& u(-x^+\gg L_+\, ,|x^-|\lesssim L_-)&=u^{in}(x^-)\,  ,
\label{FE070}\\
& w(|x^+|\lesssim L_+\, , -x^-\gg L_-)&=w^{out}(x^+)\, ,
\label{FE080}\\
& u(x^+\gg L_+\, ,|x^-|\lesssim L_-)&=u^{out}(x^-)\, .
\label{FE088}
\end{eqnarray}
Outside all these regions, including the $\Omega$ region, either the field $u$ or the field $w$, or both 
are exponentially small or vanish.

The scattering problem can be formulated now as follows: to find the outgoing waves 
$w^{out}(x^+)$ and  $u^{out}(x^-)$, knowing the incoming waves $w^{in}(x^+)$ and  $u^{in}(x^-)$.

\subsection{Perturbative solution of  1st order} \label{integr}

Equations ~(\ref{FE030}) and~(\ref{FE040}) can be rewritten in the following integral form that takes into account  
the initial conditions~(\ref{FE060}) and~(\ref{FE070}):
\begin{equation}\label{FE090}
u+\frac{\alpha}{2}uQ^\prime(uw)=u^{in}-\frac{\alpha}{2}\int\limits_{-\infty}^{x^+}\partial_-[wQ^\prime(uw)]\, dx^+\, ,
\end{equation}
\begin{equation}\label{FE100}
w+\frac{\alpha}{2}wQ^\prime(uw)=w^{in}-\frac{\alpha}{2}\int\limits_{+\infty}^{x^-}\partial_+[uQ^\prime(uw)]\, dx^-\, .
\end{equation}

In view of Eqs.(\ref{FE080}) and (\ref{FE088})
we can formally send $x^\pm$ to $\pm\infty$ in Eqs.~(\ref{FE090}) and~(\ref{FE100}) and  obtain the following scattering relationships:
\begin{equation}\label{FE110}
u^{out}=u^{in}-\frac{\alpha}{2}\int\limits_{-\infty}^{+\infty}\partial_-[wQ^\prime(uw)]\, dx^+\, ,
\end{equation}
\begin{equation}\label{FE120}
w^{out}=w^{in}+\frac{\alpha}{2}\int\limits_{-\infty}^{\infty}\partial_+[uQ^\prime(uw)]\, dx^-\, .
\end{equation}

For sufficiently small $uw$ we are able to treat Eqs.~(\ref{FE110})  and~(\ref{FE120}) perturbatively, 
and to substitute into the integrands $u^{in}$ and $w^{in}$ instead of $u$ and $w$, respectively. Then we obtain:
\begin{equation}\label{FE130}
u^{out}=u^{in}-\frac{\alpha}{2}\left[\int\limits_{-\infty}^{+\infty}(w^{in})^2Q^{\prime\prime}\left(u^{in}w^{in}\right)
\, dx^+\right] \partial_-u^{in}\, ,
\end{equation}
\begin{equation}\label{FE140}
w^{out}=w^{in}+\frac{\alpha}{2}\left[\int\limits_{-\infty}^{\infty}(u^{in})^2Q^{\prime\prime}\left(u^{in}w^{in}\right)\, dx^-\right]\partial_+w^{in}\, .
\end{equation}
For this perturbation procedure to be  be relevant, the second terms in the right hand sides of these  equations 
must  be  much 
less than the first terms. If we take for this estimation only leading term in $Q(\zeta\to 0)\propto \zeta^2$, 
then we obtain the strong
 inequality~(\ref{FE052}).
 
The leading order of $Q(\zeta)\propto \zeta^2$, 
caused by 4-photon interaction, leads  to the following delays 
 \begin{equation}
\frac{\alpha}{2}Q^{\prime\prime}\left(0\right)\int\limits_{-\infty}^{+\infty}(w^{in})^2
\, dx^+\, ,
\end{equation}
and
\begin{equation}
\frac{\alpha}{2}Q^{\prime\prime}\left(0\right)\int\limits_{-\infty}^{\infty}(u^{in})^2\, dx^-\,.
\end{equation}
of  the  $u^{out}$ and $w^{out}$ pulses and  with respect to $u^{in}$ 
and $w^{in}$, respectively.

This corresponds to the phase shift between the interacting waves discussed in Refs. \cite{FP19, AF07}.
The strong inequality~(\ref{FE052}) ensures that the delays are much shorter than $L_-$ and $L_+$, respectively. 
These delays do not correspond  to harmonic generation, although they determine  the applicability of our approach. 
The next, subleading orders of $Q(\zeta\to 0)$ cause generation of the  harmonics. 

 Our approach uses the following set of small parameters: 
$\alpha$, $u_0w_0$, $\alpha w_0^2 L_+/L_-$, and $\alpha u_0^2 L_-/L_+$.
Here we address the question whether 
 the next order corrections to Eqs.~(\ref{FE130})
 and~(\ref{FE140}), as they would appear after the next iterations of Eqs.~(\ref{FE090}) and~(\ref{FE100}), 
 or the subleading terms in $Q$ are more important for harmonic generation.
 To answer this question we use the structure of $\bar{\cal L}$, given by Eqs.~(\ref{grangina1}), 
 (\ref{Qdef}) and~(\ref{Qser}), and
  Eqs.~(\ref{FE090}) and~(\ref{FE100}) as the  equations of our perturbation theory starting from zero order: 
  $u=u^{in}$ and $w=w^{in}$. 
  The combined result of the iterations of Eqs.~(\ref{FE090}) and~(\ref{FE100}) 
  and of  retaining the higher order terms in the expansion of  the function $Q(u,w)$ can be written symbolically as
\begin{eqnarray}
u^{out}&=&u^{in} +\sum\limits_{\ell=1}^\infty\sum\limits_{m=0}^\infty \hat{U}_{\ell m} w(\alpha uw)^\ell (uw)^m\Biggr|^{in}\, ,
\label{FE150}\\
w^{out}&=&w^{in} +\sum\limits_{\ell=1}^\infty\sum\limits_{m=0}^\infty \hat{W}_{\ell m} u(\alpha uw)^\ell (uw)^m\Biggr|^{in}\, ,
\label{FE160}
\end{eqnarray}
where $\hat{U}_{\ell m}$ and $\hat{W}_{\ell m}$ are linear integro-differential operators 
that do not contain any small or large parameters. 
They act implicitly  on all arguments $x^\pm$ to the right of them separately.  
Here we use the expansion~(\ref{Qser}) of $Q(\zeta)$. 

Due to  the conservation of  the kinematic momentum, the harmonic with the number $n$ of the wave $u$ 
 will be generated by the terms proportional to $u^n$, $u^{n+2}$, etc. The largest term of this type corresponds 
 to the term in Eq.~(\ref{FE150}) with $(\ell,m)=(1,n-1)$. All other terms contributing to the same $n$-harmonic 
 will have higher orders of $\alpha$ or/and higher orders of $uw$. Thus the leading order contribution to the amplitude 
 of  the $n$-harmonic comes from the term of the form $w(\alpha uw)u^{n-1}w^{n-1}$  
 in agreement with  Eq. ~(\ref{FE130}). The same conclusion can  be drawn for the $w$-wave.
 
 We may conclude  that the higher order terms in the expansion 
 of  the function $Q(uw)$  determine the  leading order contribution to harmonic generation, and hence
 the expressions~(\ref{FE130}) and~(\ref{FE140}) are sufficient to calculate the  intensities of the harmonics 
 in the leading order relatively to the small parameters.

\section{High Order Harmonics}
\label{harm}
\subsection{General considerations}

We consider here two plane waves of the same polarization, when the  incident spectrum corresponding 
to the wave $u^{in}$ 
is narrow enough. 
In this case we may choose $u^{in}$ in the form:
\begin{equation}\label{H010}
u^{in}(x^-)=\mbox{Re}\,  \left\{U_0(x^-)\, e^{i\omega_-x^-}\right\}\, ,
\end{equation}
where $U_0$ is a smooth function of its argument with maximum value $\bar{U}_0$, and decaying 
at least exponentially outside the region $|x^-|\lesssim L_-$.  The carrier frequency of the wave is $\omega_-$. 
We assume that $\omega_-L_-$ is 
significantly larger than one.
 We assume that $U_0$ is almost constant at the spatial-time scales of the order of $1/\omega_-$. 
 The spectral relative  half width of the wave is about
\begin{equation}
\frac{\Delta \omega_-}{\omega_-}\sim\frac{1}{\omega_-L_-}\, .
\end{equation}
Instantaneous intensity of this wave  (averaged over its carrier period) is equal to
\begin{equation}\label{H030}
I_1=\frac{m^4}{8\pi\alpha} U_0U_0^*.
\end{equation}
Here the asterisk `$^*$' denotes the complex conjugation. 
This expression is in accordance to the expressions for the energy momentum tensor~(\ref{T010})-(\ref{T040}) 
in terms of $u$ and $w$. We will skip analogous comments below.

It is obvious that the spectral width of the $n$-harmonic, $\Delta \omega_-^{(n)}$ grows with $n$. 
We calculate $\Delta \omega_-^{(n)}$ below in this section. We are able to separate the  harmonics 
with adjacent numbers, 
when $\Delta\omega_-^{(n)}$ is less than, for example, half spectral distance between the harmonics, $\omega_-$:
\begin{equation}\label{H034}
2\Delta \omega_-^{(n)}< \omega_-\,.
\end{equation}

 We will not be interested here in the detailed  form of the  harmonic spectral line for separated harmonics, 
 but only in  the total intensity in the harmonic of order $n$, integrated over its spectral line.

As we explain above,  the leading term of  the $n$-harmonic wave field, $u_{(n)}^{out}(x^-)$, 
is determined by the following expression, obtained from Eq.~(\ref{FE110}):
\begin{equation}\label{H050}
-\alpha (n+1)b_{n+1} L_+ \partial_{-} (u^{in})^n \left\langle (w^{in})^{n+1}\right\rangle\, ,
\end{equation}
where 
\begin{equation}\label{H052}
 \left\langle (w^{in})^{n+1}\right\rangle=\frac{1}{2L_+}\int\limits_{-\infty}^{\infty}
\left[w^{in}\left(x^+\right)\right]^{n+1}\, dx^+\, ,
\end{equation}
and  
\begin{align}\label{H060}
&
\partial_{-} (u^{in})^n =\partial_{-}\left[\frac{1}{2}\left(U_0 e^{i\omega_-x^-}+U_0^* e^{-i\omega_-x^-}\right)\right]^n
\nonumber \\ &
=\frac{1}{2^{n-1}}\partial_{-}\mbox{~Re~}\left(U_0^n e^{in\omega_-x^-}\right)+\dots
\end{align}

Only the first term in Eq.(\ref{H060}) contributes  to $u_{(n)}^{out}(x^-)$, the other  terms contribute  
to the $n-2$-, $n-4$- and lower harmonics. 
 Hence we obtain:
\begin{align}\label{H070}
& u_{(n)}^{out}(x^-)=\alpha \frac{n(n+1)b_{n+1}}{2^{n-1}}\omega^- L_+  \left\langle (w^{in})^{n+1}\right\rangle
\nonumber \\  &
\times\mbox{~Im~}\left(U_0^n e^{in\omega_-x^-}\right)\, .
\end{align}
Then the leading term of  the `instantaneous' intensity $I_n^{out}(x^-)$ of the $n$-harmonic averaged over its carrier period can be expressed as:
\begin{align}\label{H080}
& I_n^{out}(x^-)=\frac{m^4\alpha}{8\pi} \frac{n^2(n+1)^2b_{n+1}^2}{2^{2(n-1)}} \nonumber \\  &
\times\left(\omega_- L_+\right)^2  \left\langle (w^{in})^{n+1}\right\rangle^2 |U_0(x^-)|^{2n}\, .
\end{align}
We recall that this intensity is integrated over 
the spectral line of the $n$-harmonic. The normalized intensity becomes 
\begin{align}\label{H090} &
\frac{I_n^{out}}{I_1^{in}}=\alpha^2 \frac{n^2(n+1)^2b_{n+1}^2}{2^{2(n-1)}} \nonumber \\  &
\times\left(\omega_- L_+\right)^2  \left\langle (w^{in})^{n+1}\right\rangle^2 |U_0(x^-)|^{2n-2}\, .
\end{align}
This expression is invariant relative to boosting.
We may rewrite the latter expression by  order of magnitude:
\begin{equation}\label{H100}
\frac{I_n^{out}}{I_1^{in}}\propto\left(\omega_- L_-\right)^2 \left[\alpha w_0^2\frac{L_+}{L_-}\right]^2 (w_0u_0)^{2n-2}\, 
\end{equation}
where  a coefficient depending only on $n$ has been dropped. 
The last two terms  in Eq.(\ref{H100}) are small parameters ensuring the validity of our approach. 
It breaks completely even on a qualitative level, when
$ \alpha w_0^3u_0 L_+\omega_-$
becomes approximately equal to unity. 
In this case, secular effects in evolution of the wave shape become very important. 
The theory presented in Refs.~\cite{Ka19,Sol19} treated the case when the latter parameter 
is approximately equal to or much higher than unity.

Equation~(\ref{H080}) is written for a  general profile of $w^{in}(x^+)$
that does not contain any additional spatial parameter  besides $L_+$. 
However, when this profile has the form similar to Eq.~(\ref{H010}) we find that 
\begin{equation}\label{H110}
w^{in}(x^+)={\rm Re}\, \left\{ W_0(x^+)\, e^{i\omega_+x^+}\right\}\, ,
\end{equation}
with $\omega_+L_+\gg1$, then $\left\langle (w^{in})^{n+1}\right\rangle$ becomes considerably less than $W_0^{n+1}$ 
for even values of $n$. It means that in this case  
harmonics with even numbers will be significantly suppressed with respect to the estimation~(\ref{H100}).

The spectrum of the $n$-harmonic is determined by the multiplier $U_0(x^-)^n \exp({in\omega_-x^-})$ 
in the expression~(\ref{H070}). If we approximate $U_0(x^-)$ in the region close to its maximum as
\begin{equation}
U_0(x^-)\approx \bar{U}_0 \left[1-\frac{1}{2}\left(\frac{x^{-}}{L_-}\right)^2\right]\, ,
\end{equation}
where $\bar{U}_0$ is a constant equal to an amplitude of the $u^{in}$-wave, 
then for sufficiently high $n$ the spectrum of the $n$-harmonic will be determined by the following $x^-$ dependent factor:
\begin{equation}\label{H120}
\exp\left[-\frac{n}{2}\left(\frac{x^{-}}{L_-}\right)^2+in\omega_-x^-\right]\, .
\end{equation}
Its Fourier transform is proportional to
\begin{equation}
\exp\left[-\frac{(\omega-n\omega_-)^2 L^{-\, 2}}{2 n}\right]\, .
\end{equation}
Hence the spectral width is 
\begin{equation}
\Delta \omega_-^{(n)}\simeq \frac{\sqrt{n}}{L_-}\, .
\end{equation}
In this case, the condition~(\ref{H034}) for harmonic spectral separation  is fulfilled, when
\begin{equation}\label{H130}
n<n_{\max}\sim \frac{\left(\omega_-L_-\right)^2}{4}\, .
\end{equation}
The latter inequality is non-restrictive. It allows us to skip consideration of the opposite case, 
when $n$ is considerably is above $n_{\max}$.

\subsection{Formulas for the $n$-harmonic generation in physical units}

We use in this section the presentations~(\ref{H010}) and~(\ref{H110}) for the $in$-waves, with the profiles given by 
\begin{eqnarray}
U_0(x^-)&=&\bar{U}_0\exp\left[-\frac{(x^-)^2}{2(L_-)^2}\right]\, ,
\label{P010}\\
W_0(x^+)&=&\bar{W}_0\exp\left[-\frac{(x^+)^2}{2(L_+)^2}\right]\,.
\label{P020}
\end{eqnarray}

We introduce a relationship between the intensities of two {\em in}coming waves, 
averaged over carrier periods, and the parameters $|U_0|$ and $|W_0|$ as  follows:
\begin{eqnarray}
|\bar{U}_0|^2&=&\frac{I_u}{5.8095\times 10^{28}\mbox{~W/cm}^2}\, ,
\label{P030}\\
|\bar{W}_0|^2&=&\frac{I_w}{5.8095\times 10^{28}\mbox{~W/cm}^2}\,,
\label{P040}
\end{eqnarray}
where $I_u$ and $I_w$ are maximal intensities of the $u$ and $w$ $in$-waves, respectively. 

\begin{figure}
\centering
\includegraphics[width=8cm]{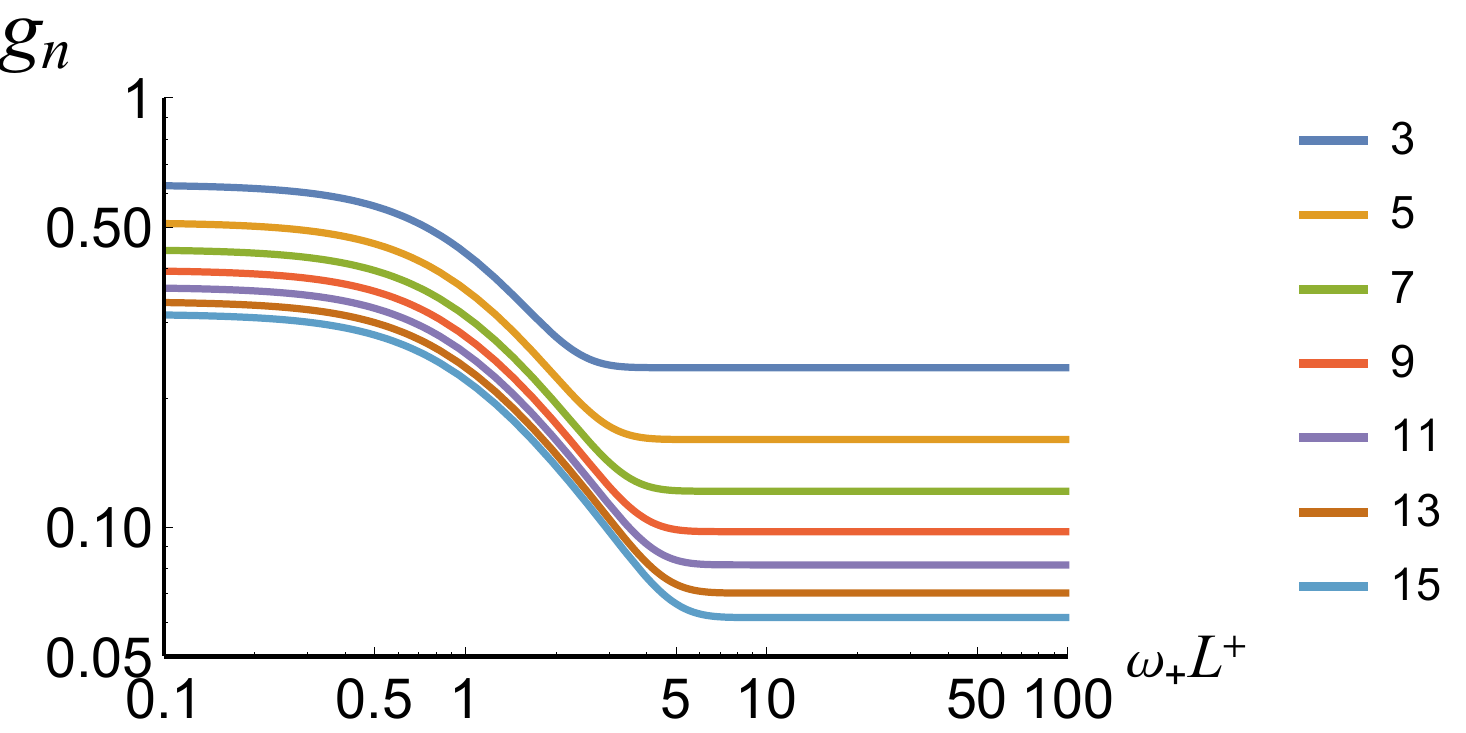}
\caption{We show  $g_n$ versus $\omega_+L_+$ for  odd values of $n=3,5,\dots, 15$.}
\label{even}
\end{figure}

\begin{figure}
\centering
\includegraphics[width=8cm]{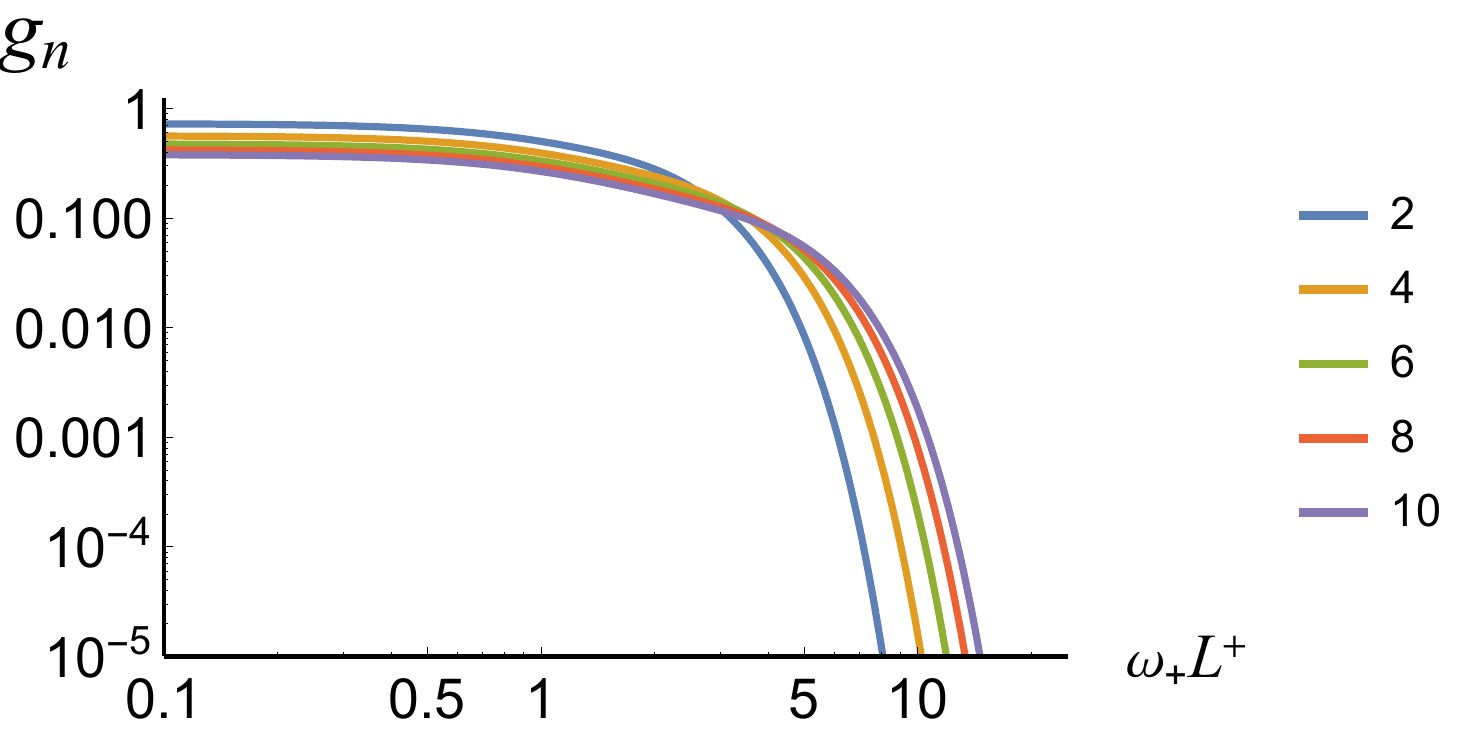}
\caption{We show $g_n$ versus $\omega_+L_+$ for even values of $n=2,4,\dots, 10$.}
\label{odd}
\end{figure}

The results described by Eqs.~(\ref{H080}) and~(\ref{H090}) contain the dimensionless parameter 
$\left\langle (w^{in})^{n+1}\right\rangle$  that depends on the $w$-pulse profile. 
We present the parameters below in the particular case where $W_0$ is a real number:
\begin{equation}\label{P050}
\left\langle (w^{in})^{n+1}\right\rangle=W_0^{n+1} g_n(\omega_+L_+)\, .
\end{equation}
We do not consider here phase envelope effects that are important for a short enough  $w$-pulse. 
We set actually a certain phase to have maximal yields for the even harmonics of the $u$-wave. 
The coefficient $g_n(\omega_+L_+)$ in Eq.~(\ref{P050}) depends only on the harmonic number 
and on the parameter $\omega_+L_+$ describing the $in$-pulse of the $w$-wave. 
The plots of $g_n$ versus $\omega_+L_+$ are shown in Figs.~\ref{even} and~\ref{odd} separately for odd and even $n$. 
For even $n$ the coefficient $g_n$ becomes equal to zero at $\omega_+L_+\to\infty$, 
whereas for odd $n$ they tend in this limit to constant values:

\begin{equation}\label{P052}
g_{2\ell-1}(\infty)=\frac{\sqrt{\pi}\,  (2\ell)!}{2^{2\ell+1}\sqrt{\ell} \, (\ell!)^2}\, .
\end{equation}

The ratio of the energy of the $n$-harmonic, ${\cal E}_n^{out}$, in the outgoing $u$-wave to the initial  
incoming energy of this wave is equal to
\begin{equation}\label{P060}
\frac{{\cal E}_n^{out}}{{\cal E}_1^{in}}=f_n\, g_{n}^2\, \left[\alpha\omega_-L_+|\bar{W}_0|^2\right]^2 \left|\bar{U}_0\bar{W}_0\right|^{2(n-1)}\, .
\end{equation}
where
\begin{equation}
f_n=\frac{n^2(n+1)^2b_{n+1}^2}{2^{2n-2}\sqrt{n}}\, .
\end{equation}
Tab.~\ref{fn} presents the values of $f_n$  and $g_n(\infty)$ for several $n$.

\begin{table}[h]
\caption{Coefficients $f_n$ and $g_{n}(\infty)$ in Eq.~(\ref{P060})}
\label{fn}
\begin{tabular}{ccc|ccc}
\toprule
$n$&$f_n$&$g_n(\infty)$&$n$&$f_n$&$g_n(\infty)$ \\
\colrule
&&&\\
2~~~&0.06568                  &~0         &~9~~ &$ 2.842\times 10^{14}$ &~0.08162      \\
3~~~&0.8580                   &~0.1599    &10~~~& $4.837\times 10^{17}$ & ~0     \\
4~~~&37.15                    &~0         &11~~~&$1.203\times10^{21}$   &~0.07017\\
5~~~&$3.973\times 10^3$       &~0.1212    &12~~~&$4.227\times 10^{24} $ &~0      \\
6~~~&$8.826\times 10^5$       &~0         &13~~~& $2.043\times 10^{28}$ &~0.06153\\
7~~~&$3.629\times 10^8 $      &~0.09753   &14~~~&$ 1.326\times10^{32}$  &~0      \\
8~~~& $2.540\times 10^{11}$   &~0         &15~~~&$1.133\times 10^{36}$  &~0.05479\\
\botrule
\end{tabular}
\end{table}

The dimensionless and Lorentz invariant factor in the brackets in Eq.~(\ref{P060}) 
is one of the small parameters of the theory. 
It is used instead of the simplified form
$\alpha  w_0^2 L_+/L_-$ introduced above  to provide a more accurate description in the limit $\omega_-L_-\gg 1$. 

The limit  $\omega_+L_+ \to 0$ in Eq.~(\ref{P060})
corresponds to the case when the $w$-wave becomes actually a  constant amplitude cross electromagnetic field. 
This fact is in accordance with the results of Refs.~\cite{Ka19,Sol19}, 
where  the generation of the second harmonic 
due to 6-photon interaction plays a main role in  the steepening of nonlinear waves 
on the constant background of  a counter-propagating cross electromagnetic field.

\section{Interaction of two electromagnetic plane waves of general polarization}
\label{APP}

\subsection{Definitions and governing equations}

We use here the same frame of reference that was introduced in Section~\ref{def}. 
Taking into account the general expression~(\ref{VP}) we may define the fields
\begin{equation}\label{C300}
u_i=\partial_-A_i\,;\quad w_i=\partial_+A_i\, ,
\end{equation}
where, and below in this section,  $i,\, j,\, \dots$ span the set $\{y,z\}$. 
The electric and magnetic field components can be expressed as 
\begin{equation}\label{C312}
E_i=\frac{u_i-w_i}{\sqrt{2}}\, ;\quad B_i=-e_{ij}\frac{u_j+w_j}{\sqrt{2}}\, .
\end{equation}
Here $e_{ij}$ is  the skew-symmetric 2-matrix with $e_{zy}=1$.
The field invariants $f$ and $g$ can be written as
\begin{eqnarray}
f&=&u_i w_i= (B^2-E^2)/2\,;
\label{C320}\\
g&=&e_{ij}u_iw_j= \left({\bf B}\cdot {\bf E}\right)/2\, .
\label{C330}
\end{eqnarray}
The Lagrangian for this problem can be written in a way similar to the case given by Eq.~(\ref{grangina1}). 
It reads
\begin{equation}\label{C340}
\mathcal{L}=-\frac{m^4}{4\pi\alpha}\left[f+\alpha Q(f,g)\right]=-\frac{m_e^4}{4\pi\alpha}\bar{\mathcal{L}}\, ,
\end{equation}
where
\begin{equation}\label{C340b}
\bar{\mathcal{L}}=f+\alpha Q(f,g)\, ,
\end{equation}
and
\begin{align}
Q(f,g)&=-\frac{1}{2\pi}{\cal M}\left(\sqrt{\sqrt{f^2+g^2}-f},\sqrt{\sqrt{f^2+g^2}+f}\right)\nonumber\\
\label{C340a}
&=\sum\limits_{m + n \geq 2} b_{m,\, n}\, f^mg^{n}\, .
\end{align}
Although $\bar{\mathcal{L}}$ and $Q$ are now functions of four field components
they can be presented in the form of functions of two arguments which are  two field invariants, $f$ and $g$. 
We note that the Lagrangian $\bar{\mathcal{L}}$ and $Q$, introduced in Section~\ref{currcons},  that 
describe the  interaction of  two electromagnetic waves of the same polarization are 
functions of the two field components
 $u$ and $w$, which were combined in one field invariant $uw$.

The field invariant $g$ is a  pseudo-scalar, hence ${\cal L}$ should be an even function of it. As a result 
all coefficients $b_{m,\, n}$ in the sum given by Eq. (\ref{C340a}) with odd $n$ vanish. 
It is possible to obtain explicit analytic expressions for the coefficients $b_{m,\, n}$. 
However the final expressions are too cumbersome for their presentation in this paper.  
We will present below their particular values, appearing in the problem for perpendicular polarizations.

In a way similar to  that used in Section~\ref{currcons}, the normalized Lagrangian~(\ref{C340b}--\ref{C340a}) 
gives the equations of the Heisenberg-Euler electrodynamics 
\begin{equation}\label{cons.curr2}
\partial_+ \left( \frac{\partial \bar{\cal L}}{\partial w_i }\right) + \partial_- \left( \frac{\partial \bar{\cal L}}{\partial u_i}\right)  = 0 .
\end{equation}
They can be written as
\begin{equation} \label{C360}
\partial_+\left[u_i\bar{\mathcal{L}}_f -e_{ij}u_j\bar{\mathcal{L}}_g\right]
+\partial_-\left[w_i\bar{\mathcal{L}}_f+e_{ij}w_j\bar{\mathcal{L}}_g\right]=0\, .
\end{equation}
As a result, using Eq.~(\ref{C340b}), we obtain 
\begin{align} 
&\partial_+\left[u_i+\alpha\left( \delta_{ij} Q_f - e_{ij} Q_g\right)\, u_j\right]\nonumber\\
\label{C362}
+&\partial_-\left[w_i+\alpha \left(\delta_{ij} Q_f+ e_{ij}Q_g\right)\, w_j\right]=0\, .
\end{align}
To get a closed system of the field equations in terms of $u_i$ and $w_i$ (instead of $A_i$) 
we may use the following direct consequence of Eq.~(\ref{C300})
\begin{equation} \label{C350}
\partial_-w_i=\partial_+u_i\, .
\end{equation}
\smallskip

The following system of the field equations, that is equivalent to Eqs.~(\ref{C362}--\ref{C350}), is more suitable for 
the perturbation theory in terms of $\alpha$: 
\begin{align}
&\partial_+u_i+\frac{1}{2}\alpha\Bigl\{\partial_+\left[\left( \delta_{ij} Q_f - e_{ij} Q_g\right)\, u_j\right]\nonumber\\
 \label{C370}
+&\partial_-\left[\left(\delta_{ij} Q_f+ e_{ij}Q_g\right)\, w_j\right]\Bigr\}=0\, ,
\end{align}
and 
\begin{align}
&\partial_-w_i+\frac{1}{2}\alpha\Bigl\{\partial_-\left[\left( \delta_{ij} Q_f + e_{ij} Q_g\right)\, w_j\right]\nonumber\\
\label{C380}
+&\partial_+\left[\left(\delta_{ij} Q_f- e_{ij}Q_g\right)\, u_j\right]\Bigr\}=0\, .
\end{align}

Integrating Eqs.~(\ref{C370}) and~(\ref{C380}) along the lines $x^{\mp}=$~const 
respectively, we obtain the following exact relationships between {\em in} and {\em out} states:
\begin{align}
u_i^{out}(x^-)&=u_i^{in}(x^-)\nonumber\\
\label{C390}
&-\frac{\alpha}{2}\int\limits_{-\infty}^\infty \partial_-\left[\left(\delta_{ij} Q_f+ e_{ij}Q_g\right)\, w_j\right]\, dx^+\, ,
\end{align}
and
\begin{align}
&w_i^{out}(x^+)=w_i^{in}(x^+)\nonumber\\
\label{C400}
&+\frac{\alpha}{2}\int\limits_{-\infty}^\infty \partial_+\left[\left(\delta_{ij} Q_f- e_{ij}Q_g\right)\, u_j\right]\, dx^-\, .
\end{align}

\subsection{Perturbative approach}

Here we consider the case of small $|f|$ and $|g|$. 
We substitute in the expressions on the right hand sides of Eqs.~(\ref{C390}) and~(\ref{C400}) 
 the unperturbed $u$ and $w$ i.e.  $u^{in}(x^-)$ and $w^{in}(x^+)$. 
Assuming  strong inequalities similar to those given by Eq. (\ref{FE052}),  we obtain 
for the ${\it out}$-wave
\begin{widetext}
\begin{equation} \label{C410}
u_i^{out}(x^-)=u_i^{in}(x^-)-\frac{\alpha}{2}\, \left(\partial_-u_k^{in}\right)\,
\int\limits_{-\infty}^\infty 
w_j^{in} w_\ell^{in}
\biggl[\delta_{ij}\delta_{k\ell}Q_{ff}+
\left(\delta_{ij}e_{k\ell}+e_{ij}\delta_{k\ell}\right)\, Q_{fg} +e_{ij}e_{k\ell}Q_{gg}\biggr]^{in}\, dx^+
\, .
\end{equation} 
\end{widetext}
The subscript $^{in}$ at the bracket means that $f$ and $g$ entering it are calculated with $u_i^{in}$ and $w_i^{in}$. 
Here the $u_i^{in}$ waves do not depend on the $x^+$ coordinate. 
Similar expressions can be written for $w_i^{out}-w_i^{in}$ waves.

\subsection{Harmonic generation by two counter-propagating waves with perpendicular polarizations}

To demonstrate how polarization can affect harmonic generation we consider here the case where the  incoming 
waves have only $u_z$ and $w_y$ components, i.e. where $u_y^{in}=w_z^{in}=0$. In this case Eq.~(\ref{C410}) becomes:
\begin{align} \label{C420} &
u_z^{out}(x^-)=u_z^{in}(x^-)-\frac{\alpha}{2}\, \left(\partial_-u_z^{in}\right) \nonumber \\&
\times\int\limits_{-\infty}^\infty Q_{gg}(0,u_z^{in}w_y^{in})\,
\left(w_y^{in}\right)^2 
\, dx^+
\, .
\end{align} 
Note that $f=0$ and $g=u_zw_y$ exactly in this case. Hence we need to know only the coefficients $b_{0,\, n}$ in the expansion~(\ref{C340a}). They are equal to:
\begin{equation} \label{C430}
b_{0,\, 2\ell}=-\frac{2^{4\ell-1}\Gamma(4\ell-2)}{\pi}\sum\limits_{\sigma=0}^{2\ell} (-1)^{\sigma+1}\frac{B_{4\ell-2\sigma}\, B_{2\sigma}}{(4\ell-2\sigma)!\, (2\sigma)!}
\end{equation}
for $\ell=1,2,\dots$.
These coefficients can be obtained from the general expansion~\cite{Dunne} of ${\cal M}({\mathfrak e},{\mathfrak b})$ in series of powers of $\mathfrak e$, and  $\mathfrak b$, setting ${\mathfrak e}={\mathfrak b}=g$.
Several leading coefficients $b_{0,\, n}$  are presented in Tab.~\ref{b02l}. We have at $\ell\to\infty$:
\begin{equation} \label{C450}
b_{0,\, 2\ell\gg 1}\simeq-\frac{8}{(4\ell)^{5/2}\sqrt{\pi}}\left(\frac{4\ell}{e\pi}\right)^{4\ell}
\end{equation}

\begin{table}[h]
\caption{Coefficients $b_{0,\, n}$ }
\label{b02l}
\begin{tabular}{c|cccc}
\toprule
&$n=2$&4&6&8 \\
\colrule
\\
$b_{0,\,n}$&$-\frac{7}{90\pi}$~~~ &~$-\frac{76}{945\pi}$~       &~$-\frac{185984}{75075\pi}$~ &~$-\frac{7007744}{16065\pi}$    \\
\botrule
\end{tabular}
\end{table}

We see that  expression~(\ref{C420}) is quite similar to what we have in Section~\ref{integr}, 
see for comparison Eq.~(\ref{FE130}). The differences are only in  the subscripts $y$ and $z$ 
and in  the appearance  of this different coefficients in the present expansion of $Q_{gg}$ in comparison 
with the coefficients in the expansion of $Q_{ff}$ (in our present terminology). 
These correspondences allow us to use our previous expressions~(\ref{H090}) and~(\ref{P060}) 
to obtain new expressions for  the intensities of  the high order harmonics for the present combinations of polarizations.

In this way we obtain the following results.  The normalized intensity of the $n$-harmonic 
(integrated over the separate spectral line of the harmonic)  becomes ($n\geq3$):
\begin{align} \label{C460} &
\frac{I_{z,\,n}^{out}}{I_{z,\,1}^{in}}=\alpha^2 \frac{n^2(n+1)^2b_{0,\, n+1}^2}{2^{2(n-1)}}
\nonumber \\ &
\times\left(\omega_- L_+\right)^2  \left\langle (w_y^{in})^{n+1}\right\rangle^2 |U_{z,\,0}(x^-)|^{2n-2}\, .
\end{align}
Here the subscripts $y$ and $z$ indicate the  direction of the electric fields of  the $u$- and $w$-waves respectively. 
The subscripts $1$ and $n$ indicate the number of the harmonic in {\em incoming} and {\em outgoing} $u_z$-waves. 
The parameters $\omega_-$ and $U_{z,\, 0}(x^-)$ describe 
the incoming $u_z$-wave in accordance with  definition~(\ref{H010}). We imply the  obvious insertion of  
the subscript $z$ into that expression. The parameters $L_+$ and $\left\langle (w_y^{in})^{n+1}\right\rangle$, 
describing the {\em incoming} $w_y$-wave are defined by Eq.~(\ref{H052}) with the obvious insertion of the $y$ 
subscript. Here, $L_+$ is the  finite length of the $w_y$-wave in the $x^+$ direction. 

For the total relative energy of  the $n$-harmonic in the {\em outgoing} $u_z$-wave we  now have
\begin{equation}\label{C470}
\frac{{\cal E}_{z,\,n}^{out}}{{\cal E}_{z,\,1}^{in}}=\bar{f}_n\, g_{z,\, n}^2\, \left[\alpha\omega_-L_+|\bar{W}_{y,\,0}|^2\right]^2 \left|\bar{U}_{z,\,0}\bar{W}_{y,\,0}\right|^{2(n-1)}\, .
\end{equation}
where
\begin{equation}\label{C471}
\bar{f}_n=\frac{n^2(n+1)^2b_{0,\,n+1}^2}{2^{2n-2}\sqrt{n}}\, ,
\end{equation}
and $\bar{U}_{z,\, 0}$,  $\bar{W}_{y,\, 0}$, and $g_{z,\, n}(\omega_+L_+)$ are defined  analogously to Eqs.~(\ref{P010}), (\ref{P020}) and~(\ref{P050}).
Plots for the functions  $g_{z,\, n}(\omega_+L_+)$ are presented in Fig.~\ref{even}, whereas Tab.~\ref{barfn} 
presents the values of $\bar{f}_n$ for several $n$. The limits  $g_{z,\, n}(\infty)$ are presented in Tab.~\ref{fn}. 
We may recall here that only the odd harmonics is generated for such a  combination of polarizations at least in 
this approximation, whereas  the suppression of the even harmonics for parallel polarization was only approximate 
even in our present order. We see also that the numerical coefficients $f_n$  are considerably larger than 
the numerical coefficient $\bar{f}_n$ with the same $n$:  compare Tabs.~\ref{fn} and~\ref{barfn}. It means that  
when the polarizations of the two waves are parallel the high harmonic generation, is much more effective than 
in the case of perpendicular polarizations.

\begin{table}[h]
\caption{Coefficients $\bar{f}_n$ in Eq.~(\ref{C470})}
\label{barfn}
\begin{tabular}{cc|cc}
\toprule
$n$&$\bar{f}_n$&$n$&$\bar{f}_n$ \\
\colrule
&&&\\
3~~~&0.005321             &11~~ &  $8.476\times 10^{13}$       \\
5~~~&1.331             &13~~~& $8.802\times10^{19}$       \\
7~~~&$7.061\times 10^3$                   &15~~~&$2.999\times 10^{26} $   \\
9~~~&$ 3.305\times 10^{8}$       &17~~~&$2.871\times 10^{33} $      \\
\botrule
\end{tabular}
\end{table}

\section{Discussion and Conclusions}
\label{concl}

We analyse the high order harmonics generation during the interaction in vacuum 
of two strong plane electromagnetic waves. The harmonics generation occurs inside the region of the waves intersection. 
The results obtained are presented in the frame of reference where the waves are counter-propagating. 

The theory presented in our paper takes into account the nonlinear polarization of the QED 
vacuum caused by the existence of a sea of virtual electron-positron pairs. 
The theory is developed within the framework of the Heisenberg-Euler electrodynamics implying 
relatively low amplitude and long-wavelength interacting waves. 

The regime of  high order harmonics generation is described by using the perturbation theory giving 
the expressions for the high order harmonics to leading order in terms of the small parameters.
 We show that the leading contributions to  the number $n$ harmonic  are obtained by expanding 
  the Heisenberg-Euler Lagrangian in power series of the electromagnetic field amplitudes and 
 not by iterations of the nonlinear contribution to the Heisenberg-Euler  Lagrangian. 
 In terms of the QED theory we can say that the amplitude of the $n$-th harmonic is determined  
 by the $2(n+1)$-photon Feynman diagram with one electron loop, but not by the combination 
 of several diagrams with fewer photon lines attached to the electron loops.
Although this statement is based on the consideration of two interacting plane waves  
we  may expect that it is  valid in a general case.  To see this, one can evaluate the impact of nonlinear currents 
entering the electromagnetic field equations of the Heisenberg-Euler electrodynamics, 
using only unperturbed electric $\mathbf{E}^{in}(\mathbf{x},t)$ and magnetic $\mathbf{B}^{in}(\mathbf{x},t)$ fields.  
As a result, the harmonic intensities can be calculated as the electromagnetic radiation emitted 
by these nonlinear currents. The unperturbed  $\mathbf{E}^{in}(\mathbf{x},t)$ and $\mathbf{B}^{in}(\mathbf{x},t)$ fields 
can be obtained by solving the linear Maxwell equations. The emitted electromagnetic fields,   
$\mathbf{E}^{out}(\mathbf{x},t)$ and $\mathbf{B}^{out}(\mathbf{x},t)$, can also  be found by solving 
the linear Maxwell equation 
 with an external source produced by the nonlinear currents. 
 This approach gives the leading order intensity of all harmonics for any polarization of the $in$-waves (see also 
 Refs. \cite{P05,Nar07,Fed07,KS15}).

From the expressions obtained above it follows that to generate one  quantum of the 5-th number harmonic 
in the collision of two counter-propagating laser beams focused in the $1 \mu$m focus spot
a laser intensity approximately equal to $5\times 10^{26}$~W/cm$^2$ is needed. 
Here we assume a 1$\mu$m wavelength laser beam  with 
 30~fs duration. To have the same photon amount  for the 3-rd harmonic an  intensity approximately  equal to 
 $3.3\times 10^{25}$~W/cm$^2$ is required. This estimate  corresponds to  interacting waves with  parallel 
 polarizations. The  interaction of perpendicular polarized pulses  is significantly less efficient 
 in generating  high order harmonics.

\begin{acknowledgments}

This work was supported by the project High Field Initiative (CZ.02.1.01/0.0/0.0/15 003/0000449) 
from the  European Regional Development Fund.

\end{acknowledgments}

\end{document}